 \documentclass[preprint,12pt]{elsarticle}

\usepackage{amssymb}
\usepackage{graphicx}

\newcounter{theorem}
\newtheorem{theorem}{Theorem}
\newcounter{corollary}

\newcounter{example}

\newcounter{proposition}
\newtheorem{proposition}{Proposition}

\begin{document}

\begin{frontmatter}



\title{B\'ezier developable surfaces}


\author{L. Fern\'andez-Jambrina} \address{Departamento de Matem\'atica e
Inform\'atica Aplicadas a las Ingenier\'{\i}as Civil y Naval\\
Universidad Polit\'ecnica de Madrid\\
Arco de la Victoria 4\\
E-28040-Madrid, Spain}

\begin{abstract}
In this paper we address the issue of designing developable surfaces
with B\'ezier patches.  We show that developable surfaces with a
polynomial edge of regression are the set of developable surfaces
which can be constructed with Aumann's algorithm.  We also obtain the set
of polynomial developable surfaces which can be constructed using general
polynomial curves.  The conclusions can be extended to spline surfaces
as well.
\end{abstract}

\end{frontmatter}

\section{Introduction}
\label{intro}

Developable surfaces are intrinsically plane surfaces, that is, they
are isometric to regions of the plane \cite{struik}.  Hence, lenghts,
angles and areas of the plane are preserved on transforming them into 
developable surfaces.  For this reason, they are much appreciated in
the industry.  Naval industry, for instance, uses plane sheets of
steel and it is less expensive to fold them than to modify their
curvature by other procedures involving application of heat.  Textile
industry uses plane sheets of cloth. And in architecture there are 
constructions based on developable surfaces \cite{architecture}. In 
fact, small deviations from developability can be accepted depending 
on the material \cite{arribas}, allowing the normal to a ruled surface to 
vary slightly along each ruling.

However, it is not easy to deal with developable surfaces within the
framework of CAGD, since the developability condition for a NURBS surface is
non-linear in the vertices of its control net. 

There have been several attempts to cope with this problem from 
different points of view.

One approach consists in restricting to classes of surfaces and their
boundary curves.  For instance, in \cite{lang} the developability
condition is solved for low degrees of rational curves.  In
\cite{aumann0} the boundary curves are taken to be parallel and of
degree three and four.  Interpolating developable surfaces are
constructed for them which are free of singularities.  This procedure
is extended to other degrees in \cite{frey}.  In \cite{maekawa}
B\'ezier such developable surfaces are linked with class of
differentiability $C^2$.

Projective geometry provides an elegant approach to this problem
\cite{ravani}.  The dual space is the set of planes in space and one
can view a developable surface as the one-parameter family of the
tangent planes to its rulings.  The developability condition is simple
in this framework \cite{pottmann-farin}.  This has motivated algorithms
for constructing developable surfaces based on their tangent planes
\cite{wallner}.

Another approach is the contruction of approximately developable
surfaces which may be useful for the industry \cite{chalfant}.  One
way to accomplish this is the use of spline cones \cite{leopoldseder}.
In \cite{arribas} nearly developable  \emph{spline} surfaces are used 
for naval architecture.

Another approach is based on the de Casteljau algorithm.  In
\cite{sequin} conditions are provided for control points of a B\'ezier
surface to be developable, restricting its application to degrees two
and three and to cylindrical and conical surfaces of arbitrary degree.
Also grounded on the de Casteljau algorithm, \cite{aumann} constructs 
generic B\'ezier developable surfaces through a given curve. In order 
to solve interpolation problems, degree-elevation is applied to the 
previous family of developable surfaces in \cite{aumann1}. The 
extension to spline developable surfaces, based on the De Boor 
algorithm, is provided in \cite{leonardo-developable, 
leonardo-elevation}.

In this paper we would like to show to what extent Aumann's family of
B\'ezier developable surfaces is general.  We show that developable
surfaces with a polynomial edge of regression are the ones which can
be constructed with Aumann's algorithm.  Furthermore, we would like to
know if there are other constructions of B\'ezier developable 
surfaces, based on general B\'ezier curves, which allow free 
parameters for design. We find a family of surfaces in addition to 
the ones available with Aumann's algorithm.

The paper is organised in the following way: In Section~\ref{maths} we
review the main properties of developable surfaces in differential
geometry.  In Section~\ref{developability} we interpret the
developability condition for B\'ezier surfaces, producing an
straightforward method for exporting results from differential
geometry to CAGD. An algebraic parametric equation for the edge of regression
of a B\'ezier developable surface is obtained.  In
Section~\ref{operations} we review transformations which can be
performed on patches of developable surfaces.  Several families of
B\'ezier developable surfaces are studied in Section~\ref{family}.  In
particular, we characterise Aumann's family in terms of their edge of
regression.  Finally, we study polynomial developable surfaces using
differential equations in order to produce families of surfaces with
free parameters based on general polynomial curves.

The results in this paper are obtained for polynomial developable
surfaces.  Since spline developable surfaces are characterised in a
similar fashion \cite{leonardo-developable}, the results are valid for
the spline case too.

\section{Developable surfaces}\label{maths}

A ruled surface bounded by two curves parametrised by $c(u)$, $d(u)$
is the surface formed by the segments, named rulings, linking points
with the same parameter $u$ on both curves.  They can be parametrised as
\begin{equation}\label{ruledsurf}
b(u,v)=(1-v)c(u)+vd(u),\quad u,v\in[0,1].\end{equation}

Reparametrisation of the curves allows for different ruled surfaces.
Hence our starting point are the parametrised curves, fixing the
parametrisation from the beginning.  A non-polynomial
reparametrisation would cast the parametrised surface out of the 
B\'ezier formalism.

The tangent plane to the ruled surface usually varies from one point 
to another along a ruling. Developable surfaces \cite{struik} are 
ruled surfaces for which the tangent plane is constant along each 
ruling. This is accomplished if the vectors $c'(u)$, $d'(u)$ and 
$\mathbf{v}(u):=d(u)-c(u)$ are coplanary for all values of $u$. 

This is easily seen, since at $v=0$ the tangent plane is spanned by 
$\mathbf{v}(u)$ and $c'(u)$ and at $v=1$ it is spanned by 
$\mathbf{v}(u)$ and $d'(u)$. If these tangent planes are the same, 
it is the same for all points on a ruling.

Hence, for developable surfaces, the vectors $c'(u)$, $\mathbf{v}(u)$ 
and $\mathbf{v}'(u)$ are linearly dependent. For non-cylindrical 
surfaces we get
\begin{equation}\label{lambdamu}
c'(u)=\lambda(u) \mathbf{v}(u)+\mu(u)\mathbf{v}'(u).\end{equation}

Accordingly, there are several types of developable surfaces:
\begin{enumerate}
\item Planar surfaces: Plane regions.

\item Cylinders: Developable surfaces with parallel rulings. 

\item Cones: Developable surfaces for which all rulings intersect at a
point named vertex.

\item Tangent surfaces: Developable surfaces spanned by the tangent lines to 
a curve, named edge of regression.
\end{enumerate}

Equation (\ref{lambdamu}) can be simplified by choosing a different 
curve on the developable surface, $r(u)=c(u)-\mu(u)\mathbf{v}(u)$,
\begin{eqnarray}
r'(u)=\left(\lambda(u)-\mu'(u)\right)\mathbf{v}(u).\label{veloedge}\end{eqnarray}

Except for the case $\lambda(u)\equiv\mu'(u)$ or $r'(u)\equiv0$, i.e.,
except for the case that $r(u)$ describes the vertex of a cone,
Equation (\ref{veloedge}) describes the tangent surface to $r(u)$, which is the edge of
regression of the developable surface.  We see that tangent surfaces
are the general case of developable surfaces.  In most of this paper
it will be the only case that we consider.

\section{B\'ezier developable patches}\label{developability}
        
The parametrisation $c(u)$ of a B\'ezier curve of degree $n$ and control polygon
$\{c_{0},\ldots,c_{n}\}$ can be contructed using the de 
Casteljau algorithm \cite{farin} in $n$ iterations,
\begin{eqnarray}\label{castel}
c^{1)}_{i}(u)&=&(1-u)c_{i}+u c_{i+1},\ i=0,\ldots,n-1 , 
\nonumber\\
c^{r)}_{i}(u)&=&(1-u)c^{r-1)}_{i}(u)+u c^{r-1)}_{i+1}(u)\ i=0,\ldots,n-r, 
\nonumber\\ 
c(u)&:=&c^{n)}_{0}(u)=(1-u)c^{n-1)}_{0}(u)+u c^{n-1)}_{1}(u).
\end{eqnarray}

The algorithm also provides the derivative of the curve as the 
difference of the two points in the 
last-but-one iteration,
\begin{eqnarray}c'(u)=n\left(c_1^{n-1)}(u)-c_0^{n-1)}(u)\right).
\label{polarder}\end{eqnarray}

\begin{figure}[h]\begin{center}
\includegraphics[height=0.25\textheight]{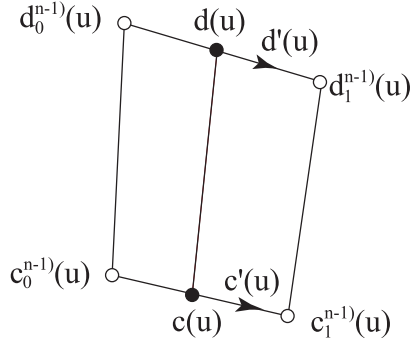}
\caption{Coplanarity condition and developability}
\label{plane}\end{center}
\end{figure}

If we consider now two B\'ezier curves $c(u)$ and $d(u)$ of degree $n$
as boundary curves of a ruled surface patch, the developability
condition of coplanarity of vectors $c'(u)$, $d'(u)$, $d(u)-c(u)$
is equivalent to the condition of coplanarity of points
$c^{n-1)}_{0}(u)$, $c^{n-1)}_{1}(u)$, $d^{n-1)}_{0}(u)$,
$d^{n-1)}_{1}(u)$.  See Fig~\ref{plane}.


This coplanarity condition can be written in terms of two functions $\Lambda(u)$, $M(u)$, such that
\[\left(1-\Lambda(u)\right)c_0^{n-1)}(u)+\Lambda(u)
c_1^{n-1)}(u)=\left(1-M(u)\right)d_0^{n-1)}(u)+M(u)d_1^{n-1)}(u),\]
except for some cases of cones.

The use of blossoms
\begin{eqnarray}
c^{1)}_{i}[u_{1}]&:=&c^{1)}_{i}(u_{1})=(1-u_{1}) c_{i}+u_{1}c_{i+1},\quad i=0,\ldots, 
n-1,\nonumber\\
c^{r)}_{i}[u_{1},\ldots,u_{r}]&:=&(1-u_{r}) 
c_{i}^{r-1)}[u_{1},\ldots,u_{r-1}]+u_{r} c_{i+1}^{r-1)}[u_{1},\ldots,u_{r-1}]
,\nonumber\\
c[u_{1},\ldots,u_{n}]&:=&c^{n)}_{0}[u_{1},\ldots,u_{n}]\ ,\quad i=0,\ldots, 
n-r,\quad r=1,\ldots,n,
\end{eqnarray}
to write the last-but-one vertices of the de Casteljau algorithm,
\[c_0^{n-1)}(u)=c[u^{<n-1>},0]\ ,\qquad
c_1^{n-1)}(u)=c[u^{<n-1>},1]\ ,\]
allow a simpler expression for the coplanarity condition 
\cite{leonardo-triangle}:

\begin{theorem}
Two B\'ezier curves $c(u)$, $d(u)$ of degree $n$ and vertices
$\{c_{0},\ldots,c_{n}\}$, $\{d_{0},\ldots,d_{n}\}$ are the boundary 
curves of a generic
developable surface patch if and only if their respective blossoms are related by
\[c[u^{<n-1>},\Lambda(u)]=d[u^{<n-1>},M(u)].\]
\end{theorem}

This construction of developable surfaces has nice features, which 
are based on the use of blossoms. Algorithms grounded on blossoms 
are compatible with this construction.

These functions $\Lambda(u)$, $M(u)$ are closely related to the ones in
(\ref{lambdamu}).  If we write that equation in terms of blossoms for
curves of degree $n$,
\begin{eqnarray*}0&=&c'(u)-\lambda(u)\mathbf{v}(u)-\mu(u)\mathbf{v}'(u)\\&=&
n\left(c[u^{<n-1>},1]-c[u^{<n-1>},0]\right)\\&-&\lambda(u)
\left(d[u^{<n>}]-c[u^{<n>}]\right)\\&-&n \mu(u)
\left(d[u^{<n-1>},1]-d[u^{<n-1>},0]\right.\nonumber
\\&-&\left.c[u^{<n-1>},1]+c[u^{<n-1>},0]\right),\end{eqnarray*} and
using the de Casteljau algorithm (\ref{castel}) for expanding the
expressions for $c(u)$ and $d(u)$, we may collect the terms,
\begin{eqnarray*}&&
c[u^{<n-1>},1]\left(n+u\lambda(u)+n\mu(u)\right) 
\\&+&c[u^{<n-1>},0]\left((1-u)\lambda(u)-n-n\mu(u)\right)\\&=& 
d[u^{<n-1>},1]\left(n\mu(u)+u\lambda(u)\right) 
\\&+&d[u^{<n-1>},0]\left((1-u)\lambda(u)-n\mu(u)\right),
\end{eqnarray*}
so that we may simplify the blossoms using affine combinations,
\begin{eqnarray*}
c\left[u^{<n-1>},\frac{n+n\mu(u)+u\lambda(u)}{\lambda(u)}\right]=
d\left[u^{<n-1>},\frac{n\mu(u)+u\lambda(u)}{\lambda(u)}\right]
.\end{eqnarray*}

If we compare this expression with Theorem 1, we get:
\begin{proposition}
A developable surface parametrised as $b(u,v)=c(u)+v\mathbf{v}(u)$, 
with $\mathbf{v}(u)=d(u)-c(u)$, 
where $c(u)$, $d(u)$ are curves of degree $n$, satisfying 
Theorem 1 for some rational 
functions $\Lambda(u)$, $M(u)$ has functions $\lambda(u)$, $\mu(u)$ 
in (\ref{lambdamu}) related by
\begin{equation}
\Lambda(u)=\frac{n(\mu(u)+1)+u\lambda(u)}{\lambda(u)},\ 
M(u)=\frac{n\mu(u)+u\lambda(u)}{\lambda(u)},
\end{equation}
\begin{equation}
\lambda(u)=\frac{n}{\Lambda(u)-M(u)},\quad 
\mu(u)=\frac{M(u)-u}{\Lambda(u)-M(u)}.
\end{equation}
\end{proposition}

In this sense we may view Theorem 1 as a way of writing the differential
equation (\ref{lambdamu}) for polynomial curves in an algebraic fashion.
This provides an interpretation for $\Lambda(u)$, $M(u)$ and it is
useful to exchange results between the general and the polynomial
case.


We may also characterise the edge of regression of a B\'ezier
developable surface in terms of the functions $\Lambda(u)$, $M(u)$.

\begin{figure}
\begin{center}
    \includegraphics[height=0.15\textheight]{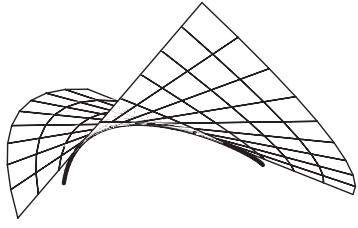}
\end{center}
\caption{Edge of regression and two sheets of a developable surface}
\label{regress}\end{figure}

The edge of regression of a developable surface is formed by the set
of points where the surface is singular, two-sheeted more precisely,
as we see in Fig.  \ref{regress}.  It is therefore of great importance
for design to keep it under control in order to produce smooth
surfaces.

We may locate it by searching the points where the parametrisation 
$b(u,v)$ of the developable surface is degenerate because its partial 
derivatives are parallel,
\[b_{u}(u,v)=(1-v) c'(u)+v d'(u)\ ,\quad
b_{v}(u,v)=d(u)-c(u)\ .\]

Since
\begin{eqnarray*}
b_{u}(u,v)&= &n(1-v)\left(c^{n-1)}_{1}(u)-c^{n-1)}_{0}(u)\right)\\&+&nv
\left(d^{n-1)}_{1}(u)-d^{n-1)}_{0}(u)\right),\\
b_{v}(u,v)&=&(1-u)\left\{d^{n-1)}_{0}(u)-c^{n-1)}_{0}(u)\right\}\\&+&
u\left\{d^{n-1)}_{1}(u)-c^{n-1)}_{1}(u)\right\},
\end{eqnarray*}
parallelism between these two vectors implies the existence of a 
factor $\alpha(u,v)$ such that $\alpha n
b_{v}=b_{u}$,
\begin{eqnarray*}&&\left(1-u+\frac{v-1}{\alpha}\right)c^{n-1)}_{0}(u)+
\left(\frac{1-v}{\alpha}+u\right)c^{n-1)}_{1}(u)\\&=&
\left(1-u+\frac{v}{\alpha}\right)d^{n-1)}_{0}(u)+
\left(u-\frac{v}{\alpha}\right)d^{n-1)}_{1}(u)\end{eqnarray*}

On the other hand the developability condition imposes another 
barycentric combination,
\[(1-\Lambda(u))c^{n-1)}_{0}(u)+\Lambda(u) c^{n-1)}_{1}(u)=
(1-M(u))d^{n-1)}_{0}(u)+M(u) d^{n-1)}_{1}(u) ,\]
which allows us to read the unknown $\alpha$
\begin{eqnarray*}\Lambda(u)=\frac{1-v}{\alpha(u,v)}+u ,\quad  
M(u)=u-\frac{v}{\alpha(u,v)},
\end{eqnarray*}
and write down the parametric equations for the edge of regression:

\begin{proposition}
A developable surface parametrised by $b(u,v)=(1-v)c(u)+vd(u)$ where 
$c(u)$, $d(u)$ are polynomial curves of degree $n$ with blossoms 
related by $c[u^{<n-1>},\Lambda(u)]=d[u^{<n-1>},M(u)]$ for some 
functions $\Lambda$, $M$ has as edge of regression the line 
given by the parametric equation
\begin{equation}\label{edge}
v=\frac{u-M(u)}{\Lambda(u)-M(u)}.\end{equation}
\end{proposition}

This is interesting, since it allows control of the position of the
edge of regression in order to keep our patch away from it, preventing
the rulings from intersecting it and ruining the surface patch.


In the constant case \cite{aumann} it implies that the edge of regression is a line 
of degree $n+1$ on the developable surface, unless $\Lambda$ equals 
$M$. 

The equation for the edge of regression is algebraic and simple enough to simplify 
the task of avoiding this singular line when designing with 
developable surfaces.

\section{Operations with B\'ezier developable surfaces}\label{operations}

The functions $\Lambda(u)$ and $M(u)$ depend on the patch for the 
developable surface. A developable surface has different functions 
$\Lambda(u)$ and $M(u)$ when considering patches with different 
boundary curves.

There are a number of simple operations that we may perform on
developable surfaces in order to modify the coordinate patch:

\begin{itemize}
    
\item Restriction of parameter $v$ to $[a,b]\subset[0,1]$: If we 
restrict the parametrisation of the developable surface through 
$v\in[a,b]$  (Fig.~\ref{restrictv}), the new boundary curves are
\[\tilde c(u)=(1-a)c(u)+a d(u),\quad \tilde d(u)=(1-b)c(u)+bd(u),\]
and their blossoms are related by Theorem~1 with new functions 
$\tilde \Lambda(u)$, $\tilde M(u)$,
\[\tilde c[u^{<n-1>},\tilde \Lambda (u)]=\tilde d[u^{<n-1>},\tilde M 
(u)],\] which in terms of the blossoms of $c(u)$ and $d(u)$,
\[c\left[u^{<n-1>},\frac{(1-a)\tilde\Lambda(u)+(b-1)\tilde M(u)}{b-a}\right]=
d\left[u^{<n-1>},\frac{b\tilde M(u)-a\tilde\Lambda(u)}{b-a}\right],\]
allows reading the functions $\Lambda(u)$, $ M(u)$ after comparing 
with the developability condition in Theorem~1,
\[\Lambda(u)=\frac{(1-a)\tilde\Lambda(u)+(b-1)\tilde M(u)}{b-a},\quad
M(u)=\frac{b\tilde M(u)-a\tilde\Lambda(u)}{b-a},\]
which provide the new functions $\tilde\Lambda(u)$, $\tilde M(u)$,
\begin{equation}\tilde\Lambda(u)=b\Lambda(u)+(1-b)M(u),\qquad 
\tilde M(u)=a\Lambda(u)+(1-a)M(u).\end{equation}
\begin{figure}[h]\begin{center}
\includegraphics[height=0.2\textheight]{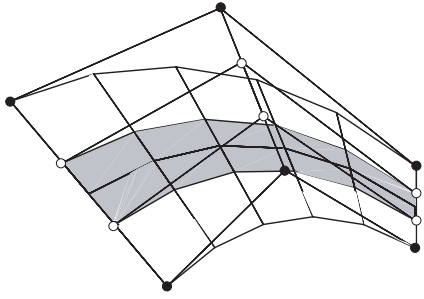}
\caption{Restriction of parameter $v$ on a developable surface\label{restrictv}}
\end{center}
\end{figure}

\item Restriction of parameter $u$ to $[a,b]\subset[0,1]$: This is 
equivalent to an affine change of parameter  $u=(1-\tilde 
u)a+\tilde ub$, $\tilde u\in [0,1]$   (Fig.~\ref{restrict}). Since blossoms are multi-affine, 
this applies directly to the relation between the functions 
$\Lambda(u)$, $M(u)$ and the new ones,
\[
\Lambda(u) =\left(1-\tilde \Lambda(u)\right)a+\tilde \Lambda(u) b\ ,\qquad
M(u) =\left(1-\tilde M(u)\right)a+\tilde M(u) b,\]
\begin{equation} \tilde \Lambda (u)
=\frac{\Lambda(u)-a}{b-a}\ , \quad \tilde M(u)
=\frac{M(u)-a}{b-a}\ .\end{equation}
\begin{figure}[h]\begin{center}
\includegraphics[height=0.2\textheight]{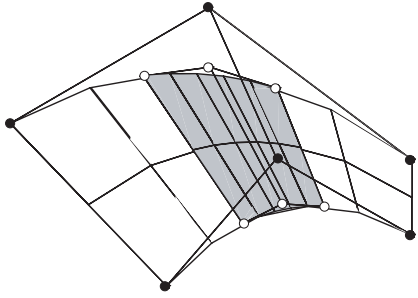}
\caption{Restriction of parameter $u$ on a developable surface}
\label{restrict}\end{center}
\end{figure}

\item Degree elevation: We know \cite{farin} that if we elevate the 
degree of a curve $c(u)$ from $n$ to $n+1$, the degree-elevated 
blossom of degree $n$ can be written in terms of the 
original blossom,
\[c^1[u_{1},\ldots,u_{n+1}]=\frac{c[u_{1},\ldots,u_{n}]+\cdots+ 
c[u_{2},\ldots,u_{n+1}]}{n+1}.\]

If we consider a developable surface spanned by two curves of degree 
$n$ $c(u)$, $d(u)$ with functions $\Lambda(u)$, $M(u)$ and elevate 
the degree of both curves to $n+1$ (See Fig.~\ref{eleva}), their blossoms have to satisfy 
the developability condition with new functions $\Lambda^1(u)$, $M^1(u)$,
\[c^1[u^{<n>},\Lambda^1(u)]=d^1[u^{<n>},M^1(u)],\]
which may be written again in terms of the original blossoms using barycentric combinations,
\begin{eqnarray*}c^1[u^{<n>},\Lambda^1(u)]&=&
\frac{nc[u^{<n-1>},\Lambda^1(u)]+c[u^{<n>}]}{n+1}\\&=&
c\left[u^{<n-1>},\frac{n\Lambda^1(u)+u}{n+1}\right].\end{eqnarray*}
\begin{figure}[h]\begin{center}
\includegraphics[height=0.2\textheight]{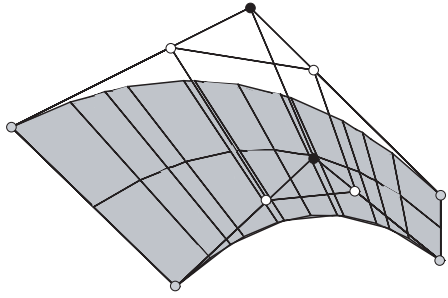}
\caption{Degree elevation of a developable surface}
\label{eleva}\end{center}
\end{figure}

The degree-elevated developability condition just states then
\[c\left[u^{<n-1>},\frac{n\Lambda^1(u)+u}{n+1}\right]=
d\left[u^{<n-1>},\frac{nM^1(u)+u}{n+1}\right]\ ,\]
and, compared with the former condition in Theorem~1, allows us to express the new 
functions  $\Lambda^1(u)$, $M^1(u)$ in terms of the previous $\Lambda(u)$, $M(u)$,
\begin{eqnarray}
 \Lambda^1(u)=\frac{(n+1)\Lambda(u)-u}{n},\qquad M^1(u)&=&\frac{(n+1)M(u)-u}{n} 
 .\label{elev}\end{eqnarray}
 
This result could be derived in a different fashion. Since in 
equation (\ref{lambdamu}) coefficients $\lambda(u)$, $\mu(u)$ are not 
altered by formal degree elevation,
\begin{eqnarray*}\lambda(u)&=&\frac{n+1}{\Lambda^1(u)-M^1(u)}=
    \frac{n}{\Lambda(u)-M(u)},\\
\mu(u)&=& 
\frac{M^1(u)-u}{\Lambda^1(u)-M^1(u)}=\frac{M(u)-u}{\Lambda(u)-M(u)},
\end{eqnarray*}
we can obtain from them the expressions for $\Lambda^1(u)$ and $M^1(u)$.

Repeated degree elevation can be also performed:

\begin{proposition}
A developable surface spanned by two curves $c(u)$, $d(u)$ of degree
$n$ with functions $\Lambda(u)$ and $M(u)$ has functions
\begin{eqnarray}\Lambda^m(u)&=&\frac{(n+m)\Lambda(u)-mu}{n},\nonumber\\ 
M^m(u)&=&\frac{(n+m)M(u)-mu}{n},\end{eqnarray}
after formally elevating the degree of the curves to $n+m$.
\end{proposition}

\item Modification of the length of the rulings: If we rewrite the parametrisation of a ruled surface interpolating 
 two curves $c(u)$, $d(u)$ in terms of the direction of the 
rulings, $\mathbf{v}(u)=d(u)-c(u)$,
we may modify the surface patch (See Fig.~\ref{enlarge}) by changing the length of the vector 
$\mathbf{v}(u)$ through a factor $h(u)$ depending on the ruling, 
$\tilde\mathbf{v}(u)=h(u)\mathbf{v}(u)$,
\[\tilde b(u,v)=c(u)+v\,h(u)\mathbf{v}(u),\]
and the edge of the surface patch moves to the curve \[\tilde 
d(u)=h(u)d(u)+\left(1-h(u)\right) c(u)\ .\]
\begin{equation}\tilde\Lambda(u)=b\Lambda(u)+(1-b)M(u),\qquad 
\tilde M(u)=a\Lambda(u)+(1-a)M(u).\end{equation}
\begin{figure}[h]\begin{center}
\includegraphics[height=0.2\textheight]{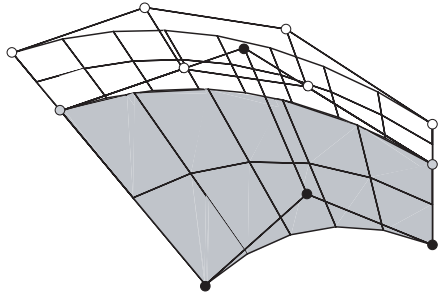}
\caption{Modification of the length of the rulings of a developable surface}
\label{enlarge}\end{center}
\end{figure}

This procedure is useful, not only for developable surfaces, but for 
ruled surfaces in general.

We compare the linear differential equation for the new patch with 
the original one,
\begin{eqnarray*}
\lambda \mathbf{v}+\mu \mathbf{v}'=c'=\tilde \lambda \tilde \mathbf{v}+
\tilde \mu\tilde \mathbf{v}'
=\left(\tilde\lambda h+\tilde\mu h'\right)\mathbf{v}+\tilde \mu h \mathbf{v}',
\end{eqnarray*}
in order to relate the old coefficients with the new ones,
\[\lambda =\tilde\lambda h+\tilde\mu h'=\frac{n}{\Lambda-M}, \quad 
\mu=\tilde \mu h=\frac{M-u}{\Lambda-M},\] \[\tilde\lambda 
=\frac{h\lambda-h'\mu}{h^2}=\frac{n+m}{\tilde\Lambda-\tilde M},\qquad \tilde \mu=\frac{\mu}{h}
=\frac{\tilde M-u}{\tilde \Lambda-\tilde M},\]
where we have assumed that $h$ is a polynomial factor of degree $m$.

This leads to relations between the functions $\Lambda$, $M$ of both 
patches,
\[
\tilde 
\Lambda(u)=u+\frac{(n+m)h^2(u)(\Lambda(u)-M(u))+(n+m)h(u)(M(u)-u)}{nh(u)-h'(u)(M(u)-u)},\]\[
\tilde M(u)=u+\frac{(n+m)h(u)(M(u)-u)}{nh(u)-h'(u)(M(u)-u)},
\]

Obviously, in the case of no enlargement, $h(u)\equiv 1$, the degree 
elevation formula (\ref{elev}) is recovered. This operation has been 
used in \cite{aumann1} for modifying developable patches.

\item Change of parameter: If we carry out the change of parameter 
$u=h(U)$, where $h$ is a polynomial function of degree $m$, the 
degree of a polynomial developable surface through a curve $c(u)$ of 
degree $n$ changes from $(n,1)$ to 
$(mn,1)$. The new differential equation for $\hat 
c(U)=c\left(h(U)\right)$, $\hat \mathbf{v}(U)=\mathbf{v}\left(h(U)\right)$,
\begin{eqnarray*}
\frac{d\hat 
c(U)}{dU}&=&h'(U)\left.\frac{dc(u)}{du}\right|_{h(U)}=h'(U)\left.\left(\lambda 
\mathbf{v}+\mu\frac{d\mathbf{v}}{du}\right)\right|_{u=h(U)}\\&=&h'(U)\lambda (h(U))
\hat\mathbf{v}(U)+\mu (h(U))\frac{d\hat\mathbf{v}(U)}{dU}\\&=&
\hat \lambda(U)\hat\mathbf{v}(U)+\hat \mu(U)\frac{d\hat\mathbf{v}(U)}{dU},
\end{eqnarray*}
allows us a relation between the coefficients,
\[\hat\lambda(U)=h'(U)\lambda\left(h(U)\right),\qquad
\hat\mu(U)=\mu\left(h(U)\right),\]
which is translated to the functions $\Lambda$, $M$,
\[\hat \Lambda(U)=\frac{
m\left(\Lambda(h(U))-h(U)\right)}{h'(U)}+U,\]\[
\hat M(U)=\frac{m\left(M(h(U))-h(U)\right)}{h'(U)}+U.\]

For instance, in the simple case $h(U)=U^{m}$,
\[\hat \Lambda(U)=\frac{\Lambda(U^{m})}{U^{m-1}},\qquad 
\hat M(U)=\frac{M(U^{m})}{U^{m-1}}.\]
\end{itemize}
\section{Some B\'ezier developable surfaces}\label{family}

\subsection{B\'ezier cylinders}

The simplest example of developable surfaces are cylinders, for which 
all the rulings are parallel to a constant vector $\mathbf{v}$.

\begin{figure}[h]\begin{center}
\includegraphics[height=0.2\textheight]{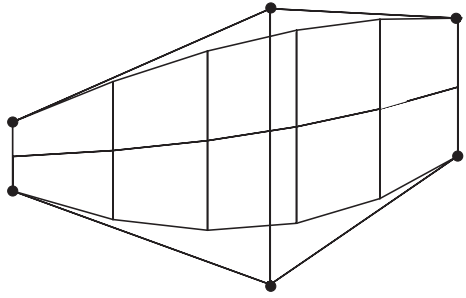}
\caption{B\'ezier cylinder}
\label{cylinder}\end{center}
\end{figure}

In the B\'ezier case, if $\{c_{0},\ldots,c_{n}\}$, 
$\{d_{0},\ldots,d_{n}\}$ are the control polygons of the boundary 
curves, this implies that all vectors $d_{i}-c_{i}$, 
$i=0,\ldots,n$ are parallel to $\mathbf{v}$ (See Fig.~\ref{cylinder}).

This implies that the differences $d^{n-1)}_{1}(u)-c^{n-1)}_{1}(u)$, 
$d^{n-1)}_{0}(u)-c^{n-1)}_{0}(u)$ must be parallel for all $u$,
\[d^{n-1)}_{1}(u)-c^{n-1)}_{1}(u)=\alpha(u)\left(d^{n-1)}_{0}(u)-c^{n-1)}_{0}(u)
\right).\]

In terms of blossoms,
\[c[u^{<n-1>},1]-\alpha(u) c[u^{<n-1>},0]=
d[u^{<n-1>},1]-\alpha(u) d[u^{<n-1>},0],\]
the property of multi-affinity allows writing it in a compact fashion,
\[c[u^{<n-1>},\Lambda(u)]=d[u^{<n-1>},\Lambda(u)], \qquad 
\Lambda(u)=\frac{\alpha(u)}{\alpha(u)-1},\]
except for the simplest case $\alpha(u)\equiv 1$, that is, 
constant $d_{i}-c_{i}=\mathbf{v}$, $i=0,\ldots,n$.

Hence, B\'ezier cylinders are developable surfaces with $\Lambda\equiv
M$.

Constructing a cylinder with direction $\mathbf{v}$ is easy,
$d(u)=c(u)+f(u)\mathbf{v}$, where $f(u)$ can be any function.  If we
require the parametrisation to be polynomial of degree $n$ in $u$, $f(u)$ is
a polynomial of degree equal or lower than $n$.  We may relate $f(u)$
with the function $\Lambda(u)$ defining the developable surface. 
Comparing
\[d(u)-c(u)=f(u)\,\mathbf{v}, \qquad d[u^{<n-1>},\Lambda(u)]=
c[u^{<n-1>},\Lambda(u)],\]
we learn that the polar form of $f$ is to fullfil
\[f[u^{<n-1>}, \Lambda(u)]\equiv 0.\]

For $f(u)=\displaystyle\sum_{i=0}^na_{i}u^i$, the previous condition
on the polarisation of $f$ reads
\[0=f[u^{<n-1>},\Lambda(u)]=a_{0}+\sum_{i=1}^n\frac{a_{i}}{n}\left((n-i)u^i+
i\Lambda(u)u^{i-1}\right),
\]
from which we can read the expression for $\Lambda(u)$,
\[\Lambda(u)=-\frac{\displaystyle\sum_{i=0}^{n-1}(n-i)a_{i}u^i}
{\displaystyle\sum_{i=1}^nia_{i}u^{i-1}}.\]

In particular, for $f(u)=(au+b)^n$, we get the constant case 
$\Lambda=-b/a$.

\subsection{B\'ezier cones}

As we have seen, cones are the subcase of developable surfaces with 
coefficients $\lambda(u)$, $\mu(u)$ related by $\lambda(u)=\mu'(u)$. 
That is, $c'(u)=\left(\mu(u)\mathbf{v}(u)\right)'$.

The simplest cone patch is the one bounded by a curve $c(u)$ with 
vertex at a point $V$. The rulings have $\mathbf{v}(u)=c(u)-V$ as 
tangent vector, but we can construct more general patches using 
$\mathbf{v}(u)=(c(u)-V)f(u)$, for which \[
\lambda(u)=-\frac{f'(u)}{f(u)},\qquad \mu(u)=\frac{1}{f(u)}.\]

If $c(u)$ is a curve of degree $n$, $c(u)-V$ is also of degree $n$ 
and except for constant $f$, modification of the patch implies 
raising its degree,\[
\Lambda(u)=\frac{f'(u)u-(n+m)(f(u)+f^2(u))}{f'(u)},\quad 
M(u)=\frac{f'(u)u-(n+m)f(u)}{f'(u)},\] if $f(u)$ is of degree $m$.

The simple case of constant $f$, for which $d(u)$ is a scaled copy of 
$c(u)$ is out of this framework.

In the linear case, $f(u)=au+b$,
\[\Lambda(u)=-\frac{(n+1)^2(au+b)^2+nau+(n+1)b}{a},\qquad M(u)=-\frac{nau+(n+1)b}{a}.\]

\subsection{Tangent surfaces}

We start with the simple case of the tangent surface to a curve 
$c(u)$ of degree $n$,
\[b(u,v)=c(u)+vf(u)c'(u),\]
which is a surface of degree $(n,1)$ provided that the factor in 
$\mathbf{v}(u)$ is linear: $f(u)=au+b$.
\begin{figure}[h]\begin{center}
\includegraphics[height=0.25\textheight]{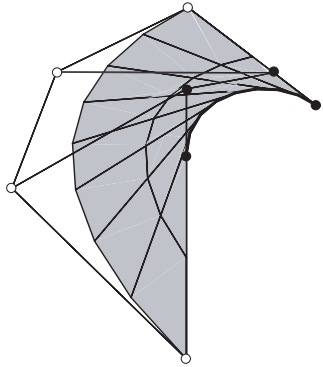}
\caption{Tangent surface to a curve}
\label{tangente}\end{center}
\end{figure}

In this case, $c'(u)=\mathbf{v}(u)/f(u)$ 
and we can read directly the functions $\lambda(u)=1/f(u)$, $\mu(u)=0$, 
and hence 
\[\Lambda(u)=u+nf(u),\qquad M(u)=u.\]

For the simplest case of $f(u)\equiv 1$ we have $\Lambda(u)=u+n$, 
$M(u)=u$. (See Fig.~\ref{tangente}).


\subsection{Case of constant $\Lambda$ and $M$}

\begin{figure}[h]\begin{center}
\includegraphics[height=0.2\textheight]{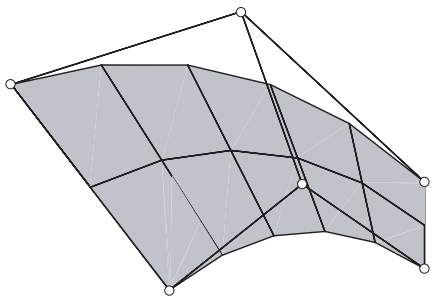}
\caption{Developable surface with constant $\Lambda$, $M$ of degree two}
\label{const}\end{center}
\end{figure}

This is the simple case that has been explored in \cite{aumann} and
extended to spline surfaces in \cite{leonardo-developable}.  This
family has the remarkable property of having linear constraints
between the vertices of the cells of the control net for the surface.
If $\{c_{0},\ldots,c_{n}\}$, $\{d_{0},\ldots,d_{n}\}$ are the control
polygons for curves $c(u)$ and $d(u)$, Theorem 1
\[
c[u^{<n-1>},\Lambda]=d[u^{<n-1>},M]
\] implies we have a set of $n$
conditions, \[(1-\Lambda)c_{i}+\Lambda c_{i+1}=(1-M)d_{i}+M d_{i+1},\
i=0,\ldots,n-1,\] which require that the cells of the surface are
planar and that the coefficients for the combinations of the vertices 
are the same for all cells (See Fig.~\ref{const} for an example).

The general solution to the linear differential equation
(\ref{lambdamu}) for these developable surfaces,
\[c'(u)=\frac{n}{\Lambda-M}\mathbf{v}(u)+\frac{M-u}{\Lambda-M}\mathbf{v}'(u),\]
can be split into the general solution to the homogeneous equation
\[n\mathbf{v}_{\mathrm{h}}(u)+(M-u)\mathbf{v}_{\mathrm{h}}'(u)=0
\Rightarrow \mathbf{v}_{\mathrm{h}}(u)=(M-u)^{n}\mathbf{w},\]
where $\mathbf{w}$ is an arbitrary constant vector, 
\[\mathbf{v}(u)=(M-u)^n\mathbf{w}+\mathbf{v}_{\mathrm{p}}(u),\]
and a particular solution of 
the whole inhomogeneous equation, $\mathbf{v}_{\mathrm{p}}(u)$, which we may 
choose of degree lower than $n$. Hence, the general case is that of 
$\mathbf{v}$ of degree $n$, except for the specific case with 
$\mathbf{w}=0$.

The edge of regression for such developable surfaces,
\[r(u)=c(u)+\frac{u-M}{\Lambda-M}(d(u)-c(u)),\]
is a polynomial curve  with velocity
\[r'(u)=\frac{n+1}{\Lambda-M}(d(u)-c(u)),\] according to 
(\ref{veloedge}). 

If $\mathbf{v}(u)$ is of degree $n$, the edge of regression is of
degree $n+1$.  This poses a paradoxical question, since the
developable surface of degree $n$ in $u$ is of degree $(n+1,1)$ in $u$
as tangent surface to its edge of regression.

In order to explain this issue, we show next that the tangent surface to a polynomial curve of degree $n+1$ can 
always be parametrised as a surface patch of degree $(n,1)$:

We start then with the tangent surface to a curve $r(u)$ of degree 
$n+1$ and define a surface patch on it, limited by two curves,
\[c(u)=r(u)+(au+b_{1})r'(u),\quad d(u)=r(u)+(au+b_{2})r'(u),\]
so that we do not change the degree of the generator of the rulings,
\[\mathbf{v}(u)=d(u)-c(u)=(b_{2}-b_{1})r'(u).\]

It is clear that we must have $a=-1/(n+1)$ in order to lower the degree of $c(u)$ 
and $d(u)$.

From the differential equation governing this patch,
\[c'(u)=\lambda(u)\mathbf{v}(u)+\mu(u)\mathbf{v}'(u)=
(1+a)r'(u)+(au+b_{1})r''(u),\]
we may read the coefficients $\lambda$, $\mu$,
\[\lambda(u)=\frac{1+a}{b_{2}-b_1},\qquad
\mu(u)=\frac{au+b_{1}}{b_{2}-b_{1}},\]
and hence the functions  for the discrete version of it,
\begin{eqnarray}
\Lambda^1(u)=\frac{\left(a(n+2)+1\right)u+(n+1)b_{2}}{a+1},\nonumber\\
M^1(u)=\frac{\left(a(n+2)+1\right)u+(n+1)b_{1}}{a+1},
\label{tangelev}\end{eqnarray}
considering that $c(u)$ and $d(u)$ are formally of degree $n+1$.

Since we want a surface patch of degree $(n,1)$, that is, with 
boundary curves of degree $n$, functions $\Lambda^1(u)$, $M^1(u)$ should 
correspond to a degree elevation (\ref{elev}). This is accomplished 
if $a=-1/(n+1)$,
\[\Lambda^1(u)=\frac{(n+1)^2b_{2}-u}{n},\qquad
M^1(u)=\frac{(n+1)^2b_{1}-u}{n},\]
which corresponds to a surface patch bounded by two curves $c(u)$, 
$d(u)$ of degree $n$ and constant functions
\[\Lambda=(n+1)b_{2},\qquad M=(n+1)b_{1}.\]

Hence we have proven that \emph{any} tangent surface to a polynomial 
curve of degree $n+1$ can be put in the form of a surface patch of 
degree $(n,1)$ with constant values of $\Lambda$, $M$.

Since, on the other hand, developable patches with constant 
$\Lambda$, $M$ have polynomial edges of regression, we have:

\begin{theorem}The set of developable surfaces with patches generated by two 
curves $c(u)$, $d(u)$ of degree $n$, with ruling generators 
$\mathbf{v}(u)=d(u)-c(u)$ also of 
degree $n$ and blossoms related by
\[c[u^{<n-1>},\Lambda]=d[u^{<n-1>},M],\] with constant $\Lambda$, $M$ 
is the set of tangent surfaces to polynomial curves of degree $n+1$.\end{theorem}

\begin{figure}[h]\begin{center}
\includegraphics[height=0.2\textheight]{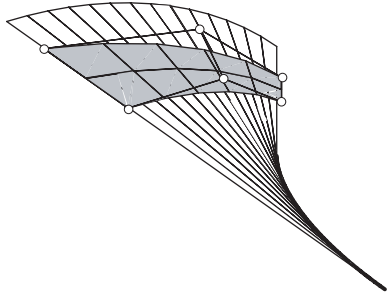}
\caption{Constant $\Lambda$, $M$ surface as a patch of the
tangent surface to a polynomial curve}
\label{regress1}\end{center}
\end{figure}

That is, Aumann's algorithm allows construction of every developable
surface with polynomial edge of regression (See Fig.~\ref{regress1}).

Besides the previous case, there are other patches
with constant functions $\Lambda$, $M$, but these are of degree
$(n+1,1)$.

For instance, for $a=-1/(n+2)$ in (\ref{tangelev}), we get
\[\Lambda=(n+2)b_{2},\qquad M=(n+2)b_{1},\]
without invoking degree elevation of curves $c(u)$, $d(u)$. 

This is a degenerated case for which $c(u)$, $d(u)$ are of degree 
$n+1$ whereas $\mathbf{v}(u)$ is of degree $n$.

Furthermore, for $a=-1/(n+m)$, we get
\[\Lambda^{2-m}(u)=\frac{(n+1)(n+m)b_{2}+(m-2)u}{n+m-1},\]\[
M^{2-m}(u)=\frac{(n+1)(n+m)b_{1}+(m-2)u}{n+m-1},\]
which correspond, after degree elevation from $n+1$ to $n+m-1$, to
\[\Lambda=(n+m)b_{2},\qquad M=(n+m)b_{1}.\]

That is, this surface patch is bounded by two curves of degree $n+1$, 
with blossoms related by constant values of $\Lambda$ and $M$, after 
formally raising their degree $m-2$ times. Since the generator of the 
rulings $\mathbf{v}(u)$ is of degree $n$, this is again a degenerated 
case.

\section{General polynomial developable surfaces}\label{diff}

According to Equation (\ref{edge}), B\'ezier developable surfaces have 
in principle a rational edge of regression. 

We have seen that we can use for design B\'ezier developable surfaces
with a polynomial edge of regression.  First, we define the surface
resorting to Aumann's construction.  Second, we modify the resulting
surface patch with the operations described in
Section~\ref{operations}.  Since Theorem 1 imposes $n$ conditions on
the control points of the boundary curves, this means that we can prescribe \cite{aumann},
for instance, just the control polygon of one of the curves,
$c_{0},\ldots,c_{n}$, and one control point of the other curve, 
$d_{0}$. Prescribing the values of $\Lambda$ and $M$ is equivalent to 
prescribe $d_{1}$ in the plane defined by $c_{0},c_{1},d_{0}$. 

However,  B\'ezier developable surfaces with a rational edge of 
regression fall out of this construction. 
%
%
%

It would be interesting to check if there are other constructions of 
polynomial developable surfaces which can be applied to B\'ezier 
curves without imposing restrictions on them and that they leave 
enough degrees of freedom for design.

With this aim in mind, we write the general solution of the linear differential equation 
(\ref{lambdamu}) in a convenient form. Since it is linear, we may split 
it as 
\begin{equation}\label{generalsol}
\mathbf{v}(u)=\mathbf{v}_{\mathrm{h}}(u)+\mathbf{v}_{\mathrm{p}}(u),\end{equation}
where $\mathbf{v}_{\mathrm{h}}(u)$ is the general solution of the 
homogeneous equation and $\mathbf{v}_{\mathrm{p}}(u)$ is one 
particular solution of the inhomogeneous equation.

The general solution of the homogeneous equation $0=\lambda 
\mathbf{v}_{\mathrm{h}}(u)+\mu\mathbf{v}_{\mathrm{h}}'(u)$ is written in 
the form
\[\mathbf{v}_{\mathrm{h}}(u)=f(u)\,\mathbf{w},\]
in terms of an arbitrary constant vector $\mathbf{w}$ and a function 
$f(u)=1/F(u)$ such that
\[-\frac{F'(u)}{F(u)}=\frac{f'(u)}{f(u)}=\frac{\mathbf{v}_{\mathrm{h}}'(u)}{\mathbf{v}_h(u)}=-
\frac{\lambda(u)}{\mu(u)},\]
and we may factor $\lambda(u)$, $\mu(u)$ by introducing another function 
$g$,
\[\lambda(u) =g(u)F'(u),\qquad \mu(u)=g(u)F(u).\]

The method of variation of constants suggests looking for a particular 
solution $\mathbf{v}_{p}(u)=f(u)\mathbf{w}(u)$, for which the 
differential equation determines $\mathbf{w'}(u)=c'(u)/g(u)$, so that 
the general solution (\ref{generalsol}) is written as
\begin{equation}
\mathbf{v}(u)=f(u)\mathbf{w}+f(u)\int 
\frac{c'(u)\,du}{g(u)}.\end{equation}

The main advantage of this form of writing the general solution is
that we can separate the influence of $f(u)$ and $g(u)$.

Our goal is to construct a parametrisation of degree $n$ in $u$ for a
B\'ezier developable surface with a procedure that is valid for
\emph{any} B\'ezier curve of degree equal or lower than $n$, allowing
for degrees of freedom.  That is, we neglect procedures that just
provide one developable surface for a given curve $c(u)$ and
procedures that are valid only for a restricted family of curves
$c(u)$.

In order to accomplish this goal, we first note that $f(u)$ must be a
polynomial of degree $\partial f\le n$.  Otherwise, we would need
$\mathbf{w}=0$ and we would get only one solution.  Since $\lambda(u)$
and $\mu(u)$ are rational functions, this implies that $g(u)$ is also
rational.  We define the degree $\partial g$ of a rational function
$g(u)$ as the degree of its numerator minus the degree of its
denominator.

Since the degree of $\mathbf{v}(u)$ must be equal or lower than $n$, 
we must have
\begin{equation} \partial f+\partial c-\partial g\le n.\end{equation}
    
The integrand $c'(u)/g(u)$ can be expanded in terms of a polynomial plus
several simple rational terms of the form $(u-a_{i})^{-m_{i}}$,
$m_{i}>0$.  Terms with power $m_{i}=1$ are to be avoided, since they
provide logarithmic terms after integration and we require polynomial
parametrisations.

This means that the polynomial terms must be also
avoided, since they appear after dividing $c'(u)$ and $g(u)$.
Depending on the form of $c'(u)$, the rational terms
$(u-a_{i})^{-m_{i}}$, could start on $m_{i}=1$ or not and we want a
procedure for every curve $c(u)$.

For the same reason, we cannot have powers of $(u-a_{i})$ and $(u-a_{j})$,
$i\neq j$, since depending on the form of $c'(u)$ the expansion could
have powers $m_{i}=1$.  Hence, in the expansion we are to have just
negative powers of just one term $(u-a)$.

But again this does not prevent terms of the form $1/(u-a)$, depending
on the form of $c'(u)$.  If we want the expansion to start at least
with $(u-a)^{-2}$, we need $\partial g \ge \partial c'+2$, that is, in 
terms of a polynomial $P_{q}(u)$ of degree $q$,
\[g(u)=\frac{(u-a)^m}{P_{q}(u)}, \quad m\ge 1+q+\partial c.\]

On integrating $c'(u)/g(u)$, a term $(u-a)^{1-m}$ arises which is to be 
cancelled by $f(u)$ in the expression of $\mathbf{v}(u)$ in order to 
be polynomial. Since $\partial f\le n$, we have another bound for $m$,
\begin{equation}\partial c+q+1\le m\le n+1.\end{equation}
    
This allows for several cases:
\begin{itemize}
    \item  $\partial c=n$: In this case $q=0$ and 
    $g(u)=A(u-a)^{n+1}$, $f(u)=(u-a)^n$,
\[\mathbf{v}(u)=(u-a)^n \mathbf{w}+ (u-a)^n\int 
\frac{c'(u)\,du}{A(u-a)^{n+1}}.\]

This parametrisation has $\lambda=-An$, $\mu(u)=A(u-a)$ and 
corresponds to a patch with constant $\Lambda$, $M$,
\[\Lambda =a-\frac{1}{A},\qquad M=a.\]

The first conclusion is that the only general construction for curves 
of degree $n$ is Aumann's \cite{aumann}.

\item $\partial c= n-1$, $q=0$, $m=n+1$: It is like the previous 
case, but with a curve $c(u)$ of degree $n-1$. It is hence the same 
case if we formally raise the degree of $c(u)$ to $n$. 

\item $\partial c= n-1$, $q=0$, $m=n$: In this case $g(u)=A(u-a)^{n}$,
$f(u)=(u-a)^{n-1}(u-b)$. The parametrisation 
\[\mathbf{v}(u)=(u-a)^{n-1}(u-b) \mathbf{w}+ (u-a)^{n-1}(u-b)\int 
\frac{c'(u)\,du}{A(u-a)^{n}}\]
shows clearly that it is just a patch of degree $n-1$ which has been 
modified by changing the length of the vector $\mathbf{v}(u)$ by a 
factor $(u-b)$.

Accordingly, the coefficients
\[\lambda(u)=A\frac{(n-1)b+a-nu}{(u-b)^2}, \qquad
\mu(u)=\frac{A(u-a)}{u-b},\] correspond to such deformation applied 
to an original patch with $\hat\lambda=-A(n-1)$, $\hat\mu(u)=A(u-a)$.

This is the degree-elevated patch described by Aumann in \cite{aumann1}.

\item $\partial c= n-1$, $q=1$: Again $m=n+1$ and 
$g(u)=A(u-a)^{n+1}/(u-b)$, $f(u)=(u-a)^n$,
\[\mathbf{v}(u)=(u-a)^n \mathbf{w}+ (u-a)^n\int 
\frac{(u-b)c'(u)\,du}{A(u-a)^{n+1}}.\]

The parametrisation is similar to the one in the first case, except
for the factor $(u-b)$.  Whereas in the previous case the length of
the rulings is modified, in this case it is modified the length of the
velocity of the curve.  It is equivalent to replace the original curve
by a new one with velocity $c'(u)(u-b)$. 

The edge of regression is rational in this case, 
\[r(u)=c(u)-\frac{A(u-a)}{(u-b)}\mathbf{v}(u).\]


In principle, it could be used for design, but it has a disadvantage
compared to the previous case.  Modifying the length of the rulings
does not change the global developable surface we start with.  But in
this case the new developable surface patch is not part of the
original surface.

\end{itemize}

Hence, we have shown that the polynomial developable surfaces which
can be constructed from general B\'ezier curves allowing for degrees
of freedom for design belong to either Aumann's family of B\'ezier
developable surfaces or to the latter family.


\section*{Acknowledgments}
 
This work is partially supported by the Spanish Ministerio de
Econom\'\i a y Competitividad through research grant TRA2015-67788-P.

\bibliographystyle{elsarticle-num}
\bibliography{cagd}

\end{document}